\documentclass[journal,twocolumn]{IEEEtran}

\usepackage{amsmath,amsthm,amssymb,scrextend}
\usepackage{fancyhdr}
\usepackage{relsize}
\usepackage{epstopdf} 
\usepackage{floatrow}
\usepackage[font={small}]{caption}
\usepackage{multirow}
\usepackage{balance}
\usepackage[numbers,sort&compress,square]{natbib}
\usepackage{ifpdf}
\ifCLASSINFOpdf
   \usepackage[pdftex]{graphicx}
   \graphicspath{{./Figures/pdf/}}
   \DeclareGraphicsExtensions{.pdf}
\else
   \usepackage[dvips]{graphicx}
   \graphicspath{{./Figures/eps/}}
   \DeclareGraphicsExtensions{.eps}
\fi

\begin{document}

\title{Accurately Accounting for Random Blockage \\ in Device-to-Device mmWave Networks}
\author{
\IEEEauthorblockN{Enass Hriba,\IEEEauthorrefmark{1}
Matthew C. Valenti,\IEEEauthorrefmark{1}
Kiran Venugopal,\IEEEauthorrefmark{2}
and Robert W. Heath, Jr.\IEEEauthorrefmark{2} } \\
\IEEEauthorrefmark{1}West Virginia University, Morgantown, WV, USA. \\
\IEEEauthorblockA{\IEEEauthorrefmark{2}The University of Texas, Austin, TX, USA.}
\vspace{-0.55cm}
}

\maketitle
\thispagestyle{empty}


\begin{abstract}
Millimeter-wave systems are characterized by the use of highly directional antennas and the presence of blockages, which significantly alter the path-loss and small-scale fading parameters.  The received power of each interferer depends on the direction it points and whether it is line-of-sight (LOS), non-LOS (i.e., partially blocked), or completely blocked.  While interferers that are sufficiently far away will almost certainly be completely blocked, a finite number of interferers in close proximity will be subject to random partial blockages.  Previous attempts to characterize mmWave networks have made the simplifying assumption that all interferers within some radius, called the LOS ball, are unblocked, while interferers beyond that radius are non-LOS.   However, compared to simulation results, the LOS ball assumption tends to overestimate outage.  In this paper, we present an accurate yet tractable analysis of finite mmWave networks that dispenses with the LOS ball assumption.   In the analysis, each interferer has a distribution that is selected randomly from several possibilities, each representing different blockage and directivity states.   First, the exact outage probability is found for a finite network with interferers in fixed locations.  Then, the spatially averaged outage probability is found by averaging over the interferer locations.  While the focus is on device-to-device networks, the analysis is general enough to find applications outside of the present mmWave framework.
\vspace{-0.2cm}
\end{abstract}


\section{Introduction}
Millimeter-wave (mmWave) has emerged in recent years as a viable candidate for both device-to-device  (D2D) communications as well as infrastructure-based (i.e., cellular) systems \cite{rappaport2014millimeter,Akdeniz,Bai2015,Singh,DiRenzo,Andrews17,Venugopal2015,Venugopal2016,torrieri2016}.  At mmWave frequencies, signals are prone to blocking by objects intersecting the propagation paths.   While the path loss could be high, it can be compensated through the use of highly directional antennas, which also helps to isolate interference.  Blocking can significantly impact the distribution of the small-scale fading (i.e., resulting in a non line-of-sight state) and if severe enough, cause the signal to be lost completely (i.e., resulting in an outage state) \cite{Akdeniz}.  The power of each received signal, whether it be a desired signal or an interfering signal, is thus highly dependent on the relative orientations of the transmit and receive antennas and the presence of objects blocking the paths.  Any meaningful analysis of mmWave systems must therefore account for antenna orientation and blockage, and typically these are modeled as appropriate random processes.

An effective methodology to study wireless systems in general, and mmWave systems in particular, is to embrace the tools of stochastic geometry to analyze the outage, coverage, and rate of wireless networks \cite{haenggi:2012}.  With stochastic geometry, the locations of the interferers and blockages are assumed to be drawn from an appropriate point process.  Stochastic geometry has been applied to mmWave cellular systems in \cite{Bai2015,Singh,DiRenzo,Andrews17} and mmWave D2D systems in \cite{Venugopal2015,Venugopal2016}.   


A survey of mathematical models and analytical techniques is provided in \cite{Andrews17} with a section devoted to blockage models.  Random shape theory, which is an offshoot of stochastic geometry, is applied in \cite{BaiVazeHeath} to carefully consider blockage effects.  When blocking is modeled as a random process, the probability that a link is line-of-sight (LOS), i.e., not blocked, is an exponentially decaying function of link distance.  The distance-dependent blocking probability causes significant challenges to the application of stochastic geometry.  This challenge can be overcome by making a simplifying assumption that all interferers within some radius, called the LOS ball, are unblocked, while interferers beyond that radius are non-LOS.  The LOS ball assumption has been applied to mmWave cellular in \cite{Bai2015,Singh} and D2D in \cite{Venugopal2016}.  Meanwhile, a two-ball approximation was applied to mmWave multi-tier cellular systems in \cite{DiRenzo}.  While it aids tractability, the LOS ball assumption causes a non-negligible loss in accuracy.  For instance, in \cite{Venugopal2016}, the LOS ball approximation caused the distribution of coverage to be underestimated by a few decibels.  

In this paper, we propose an analytical framework for mmWave networks that explicitly accounts for the blockage probabilities, thereby dispensing with the need for a LOS ball.  The key to the analysis is to break it into two steps.  In the first step, the interferers are placed in fixed locations and the outage probability found \emph{conditioned} on the interferers' locations.  Each interferer is characterized by a fading distribution that can take on a plurality of states, depending on the random orientation of the antennas and random blockage probabilities.  In the second step, the distribution of the outage is found by taking the spatial average of the conditional outage probability over the distribution of the interferer locations.   Simulation results confirm the accuracy of the strategy and demonstrate its superiority over the LOS ball assumption.

The focus of the paper is on D2D networks, whereby the interferers transmit with a common power in a uniformly distributed direction.  However, the analysis could be extended to the more complicated case of a cellular network, where each interferer's transmit power and direction are correlated with the location of its serving base station.  The analysis is generic enough that it could find applications outside of mmWave, such as in the area of frequency hopping \cite{Valenti2013}.

The remainder of the paper is organized as follows.  Section II gives a system model and provides a general problem formulation.  Section III derives an expression for the outage probability conditioned on the location of the interferers, and Section IV applies it to a D2D mmWave network.  Section V provides an approach for obtaining the spatially averaged outage probability.  Finally, the paper concludes in Section VI. 
 

\section{System Model}
Consider a wireless network with a reference receiver, a reference transmitter, and $K$ interferers located within some area $\mathcal A$.  While the network itself may have an infinite extent and therefore an infinite number of interferers, we assume that very distant interferers are fully attenuated and therefore do not contribute directly to the interference power (though they could contribute to the noise floor).  Only a finite number ($K$) of interferers are close enough to contribute to the interference power, though the contribution of each will depend critically on whether or not its signal is LOS or non-LOS.  Moreover, the number of interferers $K$ could itself be random.  For instance, if the interferers are drawn from a Poisson point process (PPP), then the number of interferers in $\mathcal A$ will be a Poisson variable.

Define the variable $S$ to represent the signal-to-interference and noise ratio (SINR) at the reference receiver.  Our goal is to find an expression for the \emph{outage probability} as a function of an \emph{SINR threshold} $\beta$, which is the cumulative distribution function (CDF) of $S$; i.e, $F_S(\beta)$.  The variable $S$ can be expressed as
\begin{eqnarray}
   S
   & = & 
   \frac{ Y_0 }{ c + \sum_{i=1}^K Y_i }. 
   \label{eg:S}
\end{eqnarray}
where $c$ is a constant related to the noise power, $Y_0$ is the received power of the reference transmitter, and $\{ Y_i \}, i \in \{1,...,K\},$ are the received powers of the $K$ interferers. We assume that $Y_0$ is a Gamma distributed random variable with a fixed shaping parameter $m_0$ and scale  parameter $\eta_0$.  



The value of $c$ is selected so that the signal-to-noise ratio $\mathsf{SNR}$ is the mean value of $S$ when the interference is turned off; i.e.
\begin{eqnarray}
 \mathsf{SNR}
 & = &
 \mathbb E
 \left[
   \frac{Y_0}{c}
 \right]  \; \; \Longrightarrow \; \;
 c
  = 
 \frac{ \mathbb E[ Y_0 ] }{ \mathsf{SNR} }.
 \label{eq:SNR}
\end{eqnarray}

The other $Y_i, i \in \{1, ..., K\},$ each have a distribution that depends on a variety of factors including the distance to the interferer, the relative orientations of the transmit and receive antennas, the random transmission activity (e.g., use of an Aloha-like protocol), and the blockage process.   We thus assume that each $Y_i, i \in \{1, ..., K\}$, is drawn from one of $J+1$ power distributions, each corresponding to a different state that encapsulates the blockage and directivity conditions. This is done by drawing a discrete random variable $a_i \in \{ 0,1,...,J \}$, which indicates the chosen power distribution.   Let $p_{i,j}$ represent the probability that $a_i=j$ for $i \in \{1, 2,..., K\}$ and  $j \in \{ 0,1,...,J \}$.  The probabilities $\{p_{i,j}\}$ could depend on the location $X_i$ of the $i^{th}$ interferer.  For instance, if a random blockage model is assumed, then the probabilities associated with blockage states will be functions of the distance to the interferer.  

Let $a_i = 0$ represent the specific case that the interferer is turned off (or not using the same resource as the reference transmitter). It follows that $Y_i = 0$ when $a_i = 0$, and thus the corresponding power distribution has probability density function (PDF) $
   f_{Y_i}(y|a_i=0) 
    = 
   \delta(y). 
$    Otherwise, when $a_i > 0$, we assume that the variable is Gamma distributed.  We define two functions: $m(a_i)$ which describes the shaping parameter associated with distribution $a_i$ and $\eta(a_i)$ which describes the scaling factor of distribution $a_i$.  The mean of $Y_i$ is $\mathbb E[Y_i] = m(a_i)/\eta(a_i)$. To make the notation more compact, we will use double subscripts for $m(\cdot)$ and $\eta(\cdot)$, so that $m(a_i=j) = m_{i,j}$ and $\eta(a_i=j) = \eta_{i,j}$. Due to path-loss and the orientation of the reference receiver's antenna, these functions generally depend on the location of the $i^{th}$ interferer, which we denote $X_i$.  
It follows that the PDF when $a_i=j$ is
\begin{eqnarray}
f_{Y_i}( y | a_i=j)
& = &
\frac{ \eta_{i,j}^{m_{i,j}} }{ \Gamma(m_{i,j}) } y^{ m_{i,j}-1 }e^{-\eta_{i,j} y } u(y)
\label{pdfi}
\end{eqnarray}
where $u(y)$ is the unit step function and $\Gamma(\cdot)$ is the Gamma function.

\section{Conditional Outage Probability}
\label{sec:COP}

\setcounter{equation}{10}
\begin{figure*}
\begin{eqnarray}
F_S(s)\hspace{-.28cm}&=&\hspace{-.28cm}1-e^{-\eta_0sc} \hspace{-.18cm}\sum_{\ell=0}^{m_0-1} \hspace{-.15cm}\frac{{\left(\eta_0sc\right)}^\ell}{\ell!} \hspace{-.15cm}\sum_{n=0}^\ell\binom{\ell}{t}\frac{t!}{c^{t}}\sum_{t_i \in {\mathcal{T}_t}}\prod_{i=1}^K \left[p_{i,0}\delta(t_i)+\sum_{j=1}^J\frac{p_{i,j}\eta_{i,j}^{m_{i,j}}}{\Gamma(m_{i,j}){t_i!}}\left(\eta_0s+\eta_{i,j}\right)^{-t_i-m_{i,j}}\Gamma(t_i+m_{i,j})\right].
\label{eq:CDF_final}
\end{eqnarray}
\vspace{0cm}
\hrulefill
\end{figure*}
\setcounter{equation}{3}

%



Assume that the interferers are in fixed locations.  From the theorem on total probability, the PDF of $Y_i, i \in \{1,...,K\}$, can simply be found from the weighted sum of the conditional probabilities.
\begin{eqnarray} 
f_{Y_i}(y)
&=&
\sum_{j=0}^J p_{i,j} f_{Y_i}(y|a_i=j) \nonumber \\
&=& 
p_{i,0}\delta(y)+\sum_{j=1}^J  \frac{p_{i,j}\eta_{i,j}^{m_{i,j}}}{\Gamma(m_{i,j})}y^{m_{i,j}-1} e^{-\eta_{i,j} y} u(y).\nonumber
\end{eqnarray}
The CDF of $S$ can then be found as
\begin{eqnarray} 
F_S(s)&=&P[S \leq s] 
=P\left[Y_0\leq s\left(c+\sum_{i=1}^K Y_i\right)\right] \nonumber  
\end{eqnarray}
\vspace{-0.5cm}
\begin{eqnarray}
&=& \idotsint\limits_{\mathbb{R}^K} \int_0^{s\left(c+\sum_{i=1}^K y_i\right)} f_{Y_0}(y_0)dy_0f_{\boldsymbol{Y}}(\boldsymbol{y})   d\boldsymbol{y}, 
\label{eq:pdfs1}
\end{eqnarray}
where $f_{\boldsymbol{Y}}(\boldsymbol{y}) $ is the joint PDF of $\mathbf Y = (Y_1,Y_2,...,Y_K)$, and the inner integral 
of $f_{Y_0}(y_0)$ 
is the CDF of $Y_0$ evaluated at $ s\left(c+\sum_{i=1}^K y_i\right)$. 
Substituting this CDF
into (\ref{eq:pdfs1}) leads to
\begin{eqnarray} 
\hspace{-2.5cm}F_S(s)
\hspace{-0.2cm}&=&\hspace{-0.2cm}1-e^{-\eta_0sc} \sum_{l=0}^{m_0-1}\frac{1}{l!}{\left(\eta_0{sc}\right)}^l \nonumber \\
& &
\hspace{-2cm}
\times\hspace{-0.1cm}
 \idotsint\limits_{\mathbb{R}^K}e^{-\eta_0{s\sum_{i=1}^K y_i}}{\left(1+\frac{\sum_{i=1}^K y_i}{c}\right)}^\ell f_{\boldsymbol{Y}}(\boldsymbol{y}) d\boldsymbol{y}.
\label{pdfs2}
\end{eqnarray}
Using the binomial theorem,
\begin{eqnarray} 
{\left(1+\frac{\sum_{i=1}^K y_i}{c}\right)}^\ell
=\sum_{t=0}^\ell\binom{\ell}{t}\frac{1}{c^{t}}\left(\sum_{i=1}^K y_i\right)^t,
\label{eq:binomial}
\end{eqnarray}
and a multinomial expansion,
\begin{eqnarray}
\left(\sum_{i=1}^K y_i\right)^t=t! \sum_{t_i \in {\mathcal{T}_t}} \prod_{i=1}^K \frac{{y_i}^{t_i}}{t_i!}
\label{eq:multinomial}
\end{eqnarray}
where ${\mathcal{T}_t}$ the set of all nonnegative $t_i$ that sum to $t$. Substituting (\ref{eq:binomial}) and (\ref{eq:multinomial}) into (\ref{pdfs2}) yields
\begin{eqnarray} 
F_S(s)&=&1-e^{-\eta_0sc} \sum_{\ell=0}^{m_0-1}\frac{1}{\ell!}{\left(\eta_0sc\right)}^\ell \sum_{t=0}^\ell\binom{\ell}{t}\frac{t!}{c^{t}} \nonumber \\
& &
\hspace{-1.25cm}
\times \hspace{-0.15cm}
\sum_{t_i \in {\mathcal{T}_t}}\prod_{i=1}^K \frac{1}{t_i!} \idotsint\limits_{\mathbb{R}^K} {y_i}^{t_i} e^{-\eta_0{s y_i}}f_{\boldsymbol{Y}}(\boldsymbol{y}) d\boldsymbol{y}.
\label{pdfs3}
\end{eqnarray}
Since $Y_1,Y_2,..., \text{and }Y_K$ are independent random variables,
(\ref{pdfs3}) can be rewritten as
\begin{eqnarray} 
F_S(s)&=&1-e^{-\eta_0sc} \sum_{\ell=0}^{m_0-1}\frac{1}{\ell!}{\left(\eta_0sc\right)}^\ell \sum_{t=0}^\ell\binom{\ell}{t}\frac{t!}{c^{t}}\nonumber \\
& &
\hspace{-1.25cm}
\times \hspace{-0.15cm}
 \sum_{t_i \in {\mathcal{T}_t}}\prod_{i=1}^K \frac{1}{t_i!}\int_0^{\infty} {y_i}^{t_i} e^{-\eta_0{s y_i}} f_{Y_i}(y_i)d{y_i}
\label{pdfs4}
\end{eqnarray}
where the integral is 
\begin{eqnarray}
\int_0^{\infty} {y_i}^{t_i} e^{-\eta_0{s y_i}} f_{Y_i}(y_i)d{y_i}
& = & p_{i,0}\delta(t_i)+ \nonumber \\
& & 
\hspace{-5cm}
\sum_{j=1}^J\frac{p_{i,j}\eta_{i,j}^{m_{i,j}}}{\Gamma(m_{i,j})}
 \left(\eta_0s+\eta_{i,j}\right)^{-t_i-m_{i,j}}\Gamma(t_i+m_{i,j}).
\label{eq:InnerInt2}
\end{eqnarray}
Substituting (\ref{eq:InnerInt2}) into (\ref{pdfs4}) gives the expression (\ref{eq:CDF_final}) at the top  of  the  page.

\setcounter{equation}{11}

\section{Application to mmWave}
\label{sec:mmWave}

\begin{figure}
\centering
\includegraphics[width=0.7\textwidth]{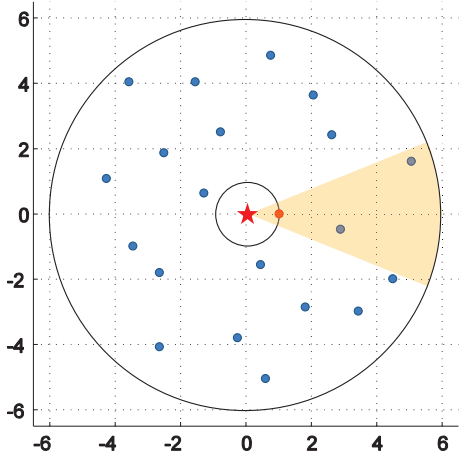}
\caption{Network Topology. The K=20 interferers are represented by the blue dots, the reference transmitter represented by the red dot, and the reference receiver represented by the red star. The yellow shaded area is the main lobe of the receiver's antenna.\vspace{-0.25cm}}
\label{fig:Network}
\end{figure}

Consider the mmWave ad hoc network shown in Fig. \ref{fig:Network}.  The reference receiver (represented by the red star) is located at the origin, while the $K$ interferers (represented by the blue dots) are located in an area $\mathcal A$, which here is assumed to be an annulus with inner radius $r_\mathsf{in}$ and outer radius $r_\mathsf{out}$.  It is assumed that a MAC protocol (such as CSMA) prevents any interference closer than $r_\mathsf{in}$ to the receiver, while the blockage is so severe at distance $r_\mathsf{out}$ that signals beyond that distance are completely attenuated.  Each interferer within $\mathcal A$ can either be unblocked, in which case its signal is LOS, or (partially) blocked, in which case its signal is non-LOS and highly (but not fully) attenuated.

The transmitter locations $X_i$ are represented by complex numbers, so that $X_i= R_i e^{j \phi_ i}$, where $R_i$ denotes the distance from the $i^{th}$ transmitter to the receiver and $\phi_i$ is the angle from $X_i$ to the receiver. The reference transmitter (represented by the red dot) is located at a distance $R_0$ from the receiver, and in this example, $R_0 = r_\mathsf{in}$. 

Assume that there are $K$ blockages in the network, and that each blockage is modeled by a disk of width $W$.  We assume that the number of blockages is the same as the number of interferers because in an mmWave ad hoc network, a main source of blockage is human bodies, and if we assume the interference is due to personal devices (e.g., wearables), then there will be approximately one interferer per person.  Assuming that the blockages are independent and uniformly distributed over the annular region, the probability that an interferer at distance $r$ from the receiver is blocked by any of the $K$ blockages is given by $p_b(r)$.  An equation and derivation for $p_b(r)$ is given in \cite{Venugopal2016}, and is incorporated herein by reference.

As in \cite{Venugopal2016,Bai2015}, we assume directional antennas that satisfy a sectorized model.  In particular, the antenna gain is $G$ inside the (half-power) beamwidth $\theta$, and $g$ outside the beamwidth.  The number of antenna elements is $N$ and the relationship between $N$, $G$, $g$, and $\theta$ is given by Table I in \cite{Venugopal2016}. We use subscripts $\mathsf{t}$ and $\mathsf{r}$ to distinguish the parameters associated with the transmitter and receiver antennas, respectively.  Thus, $N_\mathsf{r}$ is the number of elements of the receive antenna. The shaded area of Fig. \ref{fig:Network} shows the main beam of the receive antenna.  Assuming a random 2-D orientation for the interfering transmitters, the  probability that an interferer points toward the receiver is $\frac{\theta_\mathsf{t}}{2\pi}$.

We define $J=4$ transmission states corresponding to whether the interferer is or is not blocked and whether the interferer is pointing towards or away from the receiver.  In particular, we let $a_i=\{1,3\}$ when the interferer is blocked and $a_i=\{2,4\}$ when it is not, and we let $a_i = \{1,2\}$ when the interferer is pointing towards the receiver and $a_i=\{3,4\}$ when it is pointing away. Moreover, we assume an Aloha-like medium access protocol, so that the probability that the interferer transmits is $p_\mathsf{t}$.  Thus, the probability of state $a_0$, corresponding to a non-transmission state, is $(1-p_\mathsf{t})$.  It follows that the probabilities of the five states are:
\begin{eqnarray}
a_i =
\begin{cases}
0& \text{with prob. } (1-p_t) \\
1 &\text{with prob. }p_b(R_i)\frac{\theta_t}{2\pi}p_t \\
2 & \text{with prob. } (1-p_b(R_i))\frac{\theta_t}{2\pi}p_t \\
3&\text{with prob. } p_b(R_i) (1-\frac{\theta_t}{2\pi})p_t\\
4 &\text{with prob. } (1-p_b(R_i))(1-\frac{\theta_t}{2\pi})p_t. \\
\end{cases}
\label{eq:a_i}
\end{eqnarray}
Each of the above $a_i$ implies specific shaping and scale parameters for the interferer's power distribution. In particular, the value of the shaping parameter $m_{i,j}$ depends on the blockage state. When the link is blocked, i.e. when $a_i=\{1,3\}$, the shaping parameter is $m_{i,j} = m_\mathsf{N}$; otherwise $m_{i,j} = m_\mathsf{L}$, where $m_\mathsf{L}$ and $m_\mathsf{N}$ are the LOS and non-LOS shaping parameters, respectively.  

Moreover, the scaling parameter for the $i^{th}$ interferer depends on its distance $R_i$ as well as its state $a_i$, and each state could have associated with it a different antenna gain and path-loss exponent.  The $\eta_{i,j}$ parameter is given by $\eta_{i,j} = m_{i,j}/\Omega_{i,j}$
where $\Omega_{i,j}$ is the average received power given by
\begin{eqnarray}
\Omega_{i,j}
& = & 
g_r(\phi_i)g_t(a_i)R_i^{-\alpha_j},
\end{eqnarray}
the receive antenna gain is
\begin{eqnarray}
g_r(\phi_i)=\setlength{\arraycolsep}{0pt}
\renewcommand{\arraystretch}{1.2}
\left\{\begin{array}{l @{\quad} l c}
G_r &  \text{if  } |\phi_i-\phi_0| < \frac{\theta_r}{2}& \\
g_r&  \text{otherwise}&
   \end{array}\right.
\label{eq:g_r}
\end{eqnarray}
the transmit antenna gain is
\begin{eqnarray}
g_t(a_i)=\setlength{\arraycolsep}{0pt}
\renewcommand{\arraystretch}{1.2}
\left\{\begin{array}{l @{\quad} l c}
G_t &  \text{for  } a_i\in\{1,2\}& \\
g_t&  \text{for  } a_i \in \{3,4\}&
   \end{array}\right.
\end{eqnarray}
and $\alpha_j=\alpha_\mathsf{N}$ if the link is blocked and $\alpha_j=\alpha_L$ if it is not.

We assume that the reference link is LOS; i.e., $m_0=m_L$. Because the reference transmitter and reference receiver point towards one another, $\eta_0=m_0/\Omega_0$  where 
\begin{eqnarray}
{\Omega_0}=G_rG_tR_0^{-\alpha_0},
\end{eqnarray}
and $\alpha_0=\alpha_L.$

{\bf Example \#1:}  We consider as an example a network of inner radius $r_\mathsf{in} = 1$, outer radius $r_\mathsf{out} = 6$, and $K=20$ interferers.  The length of the reference link is $R_0 = r_\mathsf{in} = 1$.  The transmitters and receiver have $N_\mathsf{t}=N_\mathsf{r}=4$ antennas.  The width of each blockage is $W=1$ and we assume that there are $K$ such blockages.  The shape parameter (i.e., Nakagami-m factor) for LOS links is $m_\mathsf{L} = 4$, while that of non-LOS links is $m_\mathsf{N} = 1$ (i.e., Rayleigh fading).  The path-loss exponent for LOS links is $\alpha_\mathsf{L} = 2$, while that of non-LOS links is $\alpha_\mathsf{N} = 4$.  The probability that an interferer transmits is $p_\mathsf{t} = 0.5$, and the signal-to-noise ratio is $\mathsf{SNR}$ = 20 dB.

%

\begin{figure}
\centering
\includegraphics[width=\textwidth]{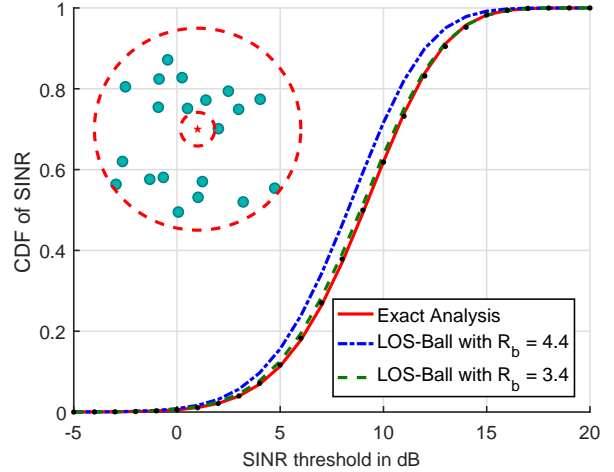}
\vspace{-0.75cm}
\caption{An example network in the upper left portion of the figure. 
The curves show the outage probability for this particular network at $\mathsf{SNR}$ = 20 dB. The black dots represent simulation results.  In addition to the exact outage probability found using the methods of this paper, the outage probability using the LOS-ball assumption is shown with two values of $R_\mathsf{LOS}$.\vspace{-0.25cm}}
\label{fig:CDF_SNR}

\end{figure}

Fig. \ref{fig:CDF_SNR} shows the outage probability for this example as a function of SINR threshold $\beta$ conditioned on the network realization shown in the left side of the figure.  The outage probability is found two ways: By using (\ref{eq:CDF_final}), which accurately accounts for the blocking probability, and by using the LOS-ball approximation, which assumes all interferers within distance $R_\mathsf{LOS}$ are LOS and those beyond that distance are non-LOS \cite{Venugopal2016}.  Two values of $R_\mathsf{LOS}$ are used.  The first, $R_\mathsf{LOS} = 4.4$ is found by matching moments; i.e., by using criterion 1 of \cite{Bai2015}.  The second, $R_\mathsf{LOS} = 3.4$ is found by selecting the value of $R_\mathsf{LOS}$ that generates an outage probability curve that most closely matches (in a mean-square error sense) the curve found by the exact analysis.  Note that finding $R_\mathsf{LOS}$ in this manner is not a sustainable solution because it requires that the exact probability be first found prior to finding the $R_\mathsf{LOS}$ that provides the best fit. Moreover, the best-fit value of $R_\mathsf{LOS}$ will change from one network realization to another.  Hence, the purpose of the curve is to give insight into the best one can do when using the LOS ball assumption, even if an ``optimal'' value of $R_\mathsf{LOS}$ were to be used.

\begin{figure}
\centering
\includegraphics[width=\textwidth]{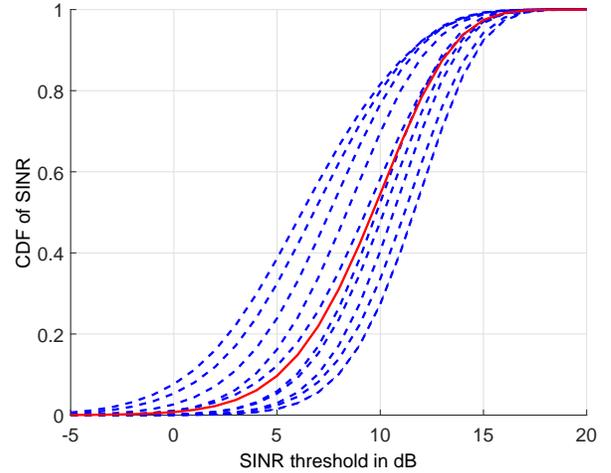}
\vspace{-0.5cm}
\caption{The outage probability conditioned on several network realization is plotted by dashed lines. The average outage probability over 100 network realization is plotted by the solid red line.\vspace{-0.5cm}}
\label{fig:FS_Avr}

\end{figure}

In addition, simulations were run to confirm the analysis, and are shown as dots on the figure.  For each value of $\mathsf{SNR}$, the simulation involved drawing 10,000 realizations of $S$, where each realization of $S$ required first drawing the necessary set of $a_i, i \in \{1, ..., L\}$, and then drawing the set of $Y_i, i \in \{0,..,K\}$.  Each dot shows the fraction of trials whose outage probability is less than the value on the ordinate.  As can be seen, the proposed analytical technique provides close agreement. Moreover, the figure shows the superiority of the exact analysis compared to the LOS-ball assumption, especially when the value $R_\mathsf{LOS}=4.4$ is used.  While $R_\mathsf{LOS}=3.4$ provides a close agreement with the simulations, finding that optimal value of $R_\mathsf{LOS}$ required the exact outage probability curves to first be found and thus its use does not simplify the overall analysis.

\section{spatially averaged outage probability}

The conditional outage probability depends significantly on the underlying network geometry.  Fig. \ref{fig:FS_Avr} shows the outage probabilities of 10 realizations of networks with $K=20$ interferers located in the annulus of inner radius $r_\mathsf{in}=1$ and outer radius $r_\mathsf{out}=6$.  The curves were found using the analytical expression (\ref{eq:CDF_final}) for a $\mathsf{SNR}$ = 15 dB.  The figure illustrates that the outage probability can vary dramatically for different network realizations.


A commonly accepted way to characterize the outage of a network of randomly located interferers is the \emph{spatially averaged} outage probability, which is found by removing the conditioning on the network geometry.   The spatially averaged outage probability could be found numerically via simulation. The simulation would involve randomly generating different network realizations, and computing the conditional outage probability of each, then averaging over many such network realizations.  As an example, the solid red line in Fig. \ref{fig:FS_Avr} shows the numerical average of the outage probability over 100 network realization.  A more sophisticated numerical technique is given in \cite{Valenti2014} which does not use simulation and will work for any arbitrary network topology.  However, for more regular shaped networks (e.g., circular, annular, or confined within a regular polygon), it is possible to get an analytical, rather than numerical, solution, as we describe in this section.

\setcounter{equation}{27}
\begin{figure*}[t]
\begin{eqnarray}
E_{\phi_i}\Bigg[Q_{t_i}^{\alpha_j}\left(\frac{g_r(\phi_i)g_t(a_i)}{T_{k+1}^{\alpha_j}}\right)T_{k+1}^{2}\hspace{-0.1cm}-\hspace{-0.05cm}Q_{t_i}^{\alpha_j}\left(\frac{g_r(\phi_i)g_t(a_i)}{T_{k}^{\alpha_j}}\right)T_{k}^{2}\Bigg]\hspace{-0.1cm} & = & \nonumber \\
& &
\hspace{-10cm}
\frac{\theta_r}{2\pi}
\left[Q_{t_i}^{\alpha_j}\left(\frac{Grg_t(a_i)}{T_{k+1}^{\alpha_j}}\right)T_{k+1}^{2}\hspace{-0.1cm}-\hspace{-0.05cm}Q_{t_i}^{\alpha_j}\left(\frac{Grg_t(a_i)}{T_{k}^{\alpha_j}}\right)T_{k}^{2}\right] 
+ 
\left(1-\frac{\theta_r}{2\pi}\right)
\left[Q_{t_i}^{\alpha_j}\left(\frac{grg_t(a_i)}{T_{k+1}^{\alpha_j}}\right)T_{k+1}^{2}\hspace{-0.1cm}-\hspace{-0.05cm}Q_{t_i}^{\alpha_j}\left(\frac{grg_t(a_i)}{T_{k}^{\alpha_j}}\right)T_{k}^{2}\right].
\vspace{-1cm}
\label{eq:Eqi}
\end{eqnarray}
\vspace{0cm}
\hrulefill
\end{figure*}
\setcounter{equation}{16}

Let $\mathbb E_{\mathbf{X}}[F_S(s)]$ denote the spatially averaged outage probability, where the expectation is with respect to $\mathbf{X} = (X_0, X_1, ..., X_K)$. From  (\ref{eq:CDF_final}) and the independence of $\{Y_i\}$, the spatially averaged outage probability can be found as follows:
\begin{eqnarray}
\mathbb E_{\mathbf X }[F_S(s)]&=& 1 - \nonumber \\ 
& & \hspace{-3cm} e^{-\eta_0sc} \sum_{l=0}^{m_0-1}\frac{1}{l!}{\left(\eta_0sc\right)}^l \sum_{n=0}^l\binom{l}{t}\frac{t!}{c^{t}}
\sum_{t_i \in {\mathcal{T}_t}}\prod_{i=1}^KE_{X_i}[\gamma_i] 
\label{eq:EFs}
\end{eqnarray}
where 
\begin{eqnarray}
\gamma_i&=& p_{i,0}\delta(t_i) + \nonumber \\
& & \hspace{-2cm}
\sum_{j=1}^J\frac{p_{i,j}}{t_i!}\frac{\eta_{i,j}^{m_{i,j}}}{\Gamma(m_{i,j})}\left(\eta_0s+\eta_{i,j}\right)^{-t_i-m_{i,j}} \Gamma(t_i+m_{i,j}). 
\label{eq:gammai}
\end{eqnarray}

If the $X_i$ are independent and uniformly distributed on an annulus, then the PDF of $R_i = |X_i|$ is $f_{R_i}(r)=\frac{2\pi r}{|A|}$ for $r_{\mathsf{in}} \leq r \leq r_{\mathsf{out}}$, and $\phi_i= \angle X_i$ is uniform over $(0,2\pi)$.  Since $R_i$ is independent of $\phi_i$, 
\begin{eqnarray}
\mathbb E_{X_i}[\gamma_i]=\mathbb E_{R_i,\phi_i}[\gamma_i]=\mathbb E_{R_i}\mathbb E_{\phi_i}[\gamma_i].
\label{eq:EXI}
\end{eqnarray}

A key challenge in finding the spatial average is that not only does the power of each interferer depend on the distance $R_i$ to the interferer, but the probabilities $p_{i,j}$ can also depend on the distance.  This makes the integral required for spatial averaging difficult, if not impossible, to evaluate in closed form.  To alleviate this issue, we divide the network $|A|$ into $L$ concentric rings and assume that for sufficiently small rings the probabilities $p_{i,j}$ are constant for all interferers in a given ring.  Let $T_0=r_{\mathsf{in}}$, $T_L=r_{\mathsf{out}}$ and $T_k=r_{\mathsf{in}}+k  \Delta r$ for $k=0,1,...,L$ where $ \Delta r= \frac{r_{\mathsf{out}}-r_{\mathsf{in}}}{L}$.
For large $L$,  we use the approximation
\begin{eqnarray} 
p_b(R_i)&\simeq& p_b\left(\frac{T_{k+1}+T_k}{2}\right)  \overset{\Delta}{=}  p_b^{(k)}  
\end{eqnarray}
for $T_k \leq R_i \leq T_{k+1}$ and $k=0,1,...,L-1$. Thus, the probability $p_{i,j}$ will be approximated by $p_{i,j}^{(k)}$, for $T_k \leq R_i \leq T_{k+1}$. 

Denoting $g_r(\phi_i)g_t(a_i)=\gamma_{i,j}$, conditioned on $T_k \leq R_i \leq T_{k+1}$ and $\phi_i$, the conditional PDF of $\Omega_{i,j}$ is
\begin{eqnarray}
f_{\Omega_{i,j}}(\omega | T_k \leq R_i \leq T_{k+1},\phi_i)&=&\frac{2 \omega^{-\frac{2+\alpha_j}{\alpha_j}}}{\alpha_j(T_{k+1}^2-T_k^2)} \gamma_{i,j}^{\frac{2}{\alpha_j}} \nonumber \\ \vspace{-1cm}
\end{eqnarray}

for $\gamma_{i,j}/T_{k+1}^{\alpha_j}  \leq \omega \leq \gamma_{i,j}/T_{k}^{\alpha_j}$.
Since
\begin{eqnarray}
 P[T_k \leq R_i \leq T_{k+1}]=\frac{\pi(T_{k+1}^2-T_k^2)}{|A|}
\end{eqnarray}
the PDF of $\Omega_{i,j}$ conditioned on $\phi_i$ is
\begin{eqnarray}
f_{\Omega_{i,j}}(\omega|\phi_i) & = & \nonumber \\
&  &
\hspace{-2.1cm}
\sum_{k=0}^{L-1}\frac{2\pi}{\alpha_j |A|} \omega^{-\frac{2+\alpha_j}{\alpha_j}} \gamma_{i,j}^{\frac{2}{\alpha_j}}\Bigg[  u\left(\omega-\frac{\gamma_{i,j}}{T_{k+1}^{\alpha_j}}\right)
- u\left(\omega-\frac{\gamma_{i,j}}{T_{k}^{\alpha_j}}\right) \Bigg]. \nonumber \\
\label{eq:pdfOmega}
\end{eqnarray}
over $\gamma_{i,j}/r_\mathsf{out}^{\alpha_j}  \leq \omega \leq \gamma_{i,j}/r_\mathsf{in}^{\alpha_j}$ and zero elsewhere.
The expectation in (\ref{eq:EXI}) can be evaluated with respect to $\Omega_{i,j}$, i.e,
\begin{eqnarray}
E_{X_i}[\gamma_i]=E_{\phi_i}E_{R_i}[\gamma_i]=E_{\phi_i}E_{\Omega_{i,j}}[\gamma_i].
\label{eq:Egamma}
\end{eqnarray}
Substituting (\ref{eq:gammai}) into (\ref{eq:Egamma}) and using the definition of $\eta_{i,j}$,
\begin{eqnarray}
E_{X_i}[\gamma_i]&=&E_{\phi_i}E_{\Omega_i}\Bigg[p_{i,0}\delta(t_i)+\sum_{j=1}^J\frac{p_{i,j}}{t_i!}\frac{\Gamma(t_i+m_{i,j})}{\Gamma(m_{i,j})}\nonumber \\
&&\times 
\frac{\Omega_{i,j}^{t_i}}{m_{i,j}^{t_i}} \left(1+\frac{\eta_0s\Omega_{i,j}}{m_{i,j}}\right)^{-t_i-m_{i,j}}\Bigg].
\label{eq:Egamma}
\end{eqnarray}
Using the PDF of conditional $\Omega_{i,j}$ in (\ref{eq:pdfOmega})
\begin{eqnarray}
\hspace{-0.9cm}&&E_{X_i}[\gamma_i]=E_{\phi_i}\Bigg[p_{i,0}\delta(t_i)+\sum_{j=1}^J  \sum_{k=0}^{L-1}\frac{2\pi\gamma_{i,j}^{\frac{2}{\alpha_j}}}{\alpha_j |A|}\frac{\Gamma(t_i+m_{i,j})}{\Gamma(m_{i,j})t_i!}\nonumber \\
\hspace{-0.9cm}&&\times \frac{p_{i,j}^{(k)}}{m_{i,j}^{t_i}}
\int_{\frac{\gamma_{i,j}}{T_{k+1}^{\alpha_j}}}^{\frac{\gamma_{i,j}}{T_{k}^{\alpha_j}}} \omega^{-\frac{2+\alpha_j}{\alpha_j}}\left(1+\frac{\eta_0s\omega}{m_{i,j}}\right)^{-t_i-m_{i,j}}\hspace{-0.3cm} \omega^{t_i}d\omega\Bigg] 
\end{eqnarray}
which evaluates to,
\begin{eqnarray}
\hspace{-0.5cm}&&p_{i,0}\delta(t_i)+\sum_{j=1}^J  \sum_{k=0}^{L-1}\frac{2\pi p_{i,j}^{(k)}}{\alpha_j |A|}\frac{\Gamma(t_i+m_{i,j})}{\Gamma(m_{i,j})t_i!}m_{i,j}^{m_{i,j}}\nonumber \\
\hspace{-0.5cm}&&\times 
(\eta_0 s)^{-m_{i,j} - t_i}E_{\phi_i}\Bigg[Q_{t_i}^{\alpha_j}\left(\frac{g_r(\phi_i)g_t(a_i)}{T_{k+1}^{\alpha_j}}\right)T_{k+1}^{2}\nonumber \\
\hspace{-0.5cm}&&-Q_{t_i}^{\alpha_j}\left(\frac{g_r(\phi_i)g_t(a_i)}{T_{k}^{\alpha_j}}\right)T_{k}^{2}\Bigg],\hspace{-1cm}
\label{eq:EPi}
\end{eqnarray}
where
\begin{eqnarray}
Q_{t_i}^{\alpha_j}\left(x\right) & = &
\frac{{}_2F_1\left(m_{i,j}+t_i,m_{i,j}+\frac{2}{\alpha_j};m_{i,j}+\frac{2}{\alpha_j}+1;-\frac{m_{i,j}}{xs\eta_0}\right)}{x^{m_{i,j}}(m_{i,j}+\frac{2}{\alpha_j})} \nonumber
\end{eqnarray}
and ${}_2F_1 (\cdot)$ is the Gauss hypergeometric function.
From (\ref{eq:g_r}), the expected value with respect to $\phi_i$ in (\ref{eq:EPi}), yields equation (\ref{eq:Eqi}) at the top  of  the  page.

%

\begin{figure}
\centering
    \includegraphics[width=\textwidth]{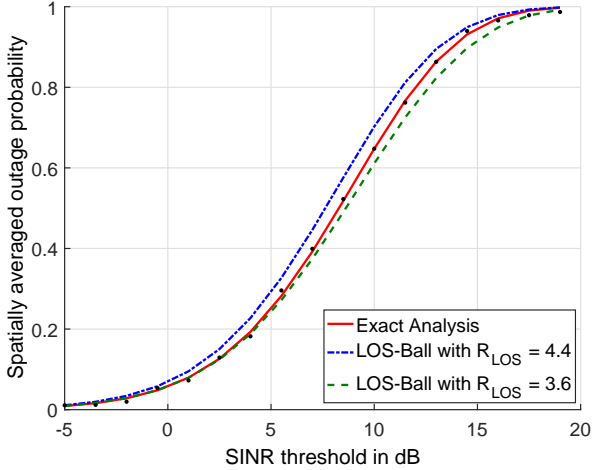}
\vspace{-0.25cm}
\caption{Spatially averaged outage probability with $K=20$ randomly located interferers and $\mathsf{SNR}$ = 20 dB. In addition to the exact values, the outage found using the LOS-ball approximation and two values of $R_\mathsf{LOS}$ are shown. The black dots represents simulation results.\vspace{-0.5cm}}
\label{fig:K}
\end{figure}

{\bf Example \#2:}  This example uses the same parameters as Example \#1 except now the $K=20$ interferers are assumed to be placed randomly within the annulus.  Rather than computing the conditional outage probability for just one network realization, Fig. \ref{fig:K} shows the spatially averaged outage probability found using (\ref{eq:EFs}) with $L=10$.  In addition to the exact analysis, the spatially averaged outage probability is found using the LOS-ball approximation with two values of $R_\mathsf{LOS}$: $R_\mathsf{LOS} = 4.4$ corresponding to criterion 1 of \cite{Bai2015} and $R_\mathsf{LOS}=3.6$, which is the value that, on average, provides the best fit to the exact outage probability.  Moreover, the dots on the figure show the spatially averaged outage probability found by averaging analytical expression for conditional outage probability (\ref{eq:CDF_final}) over 100 network realizations.  Note that the exact analysis provides close agreement with the simulation results, while both values of $R_\mathsf{LOS}$ result in a discrepancy.

\section{Conclusion}

In this paper, we found analytical expressions that exactly characterize the outage probability in wireless networks when the power of each interferer is selected at random.   The set of distributions can correspond to different blockage and directivity states, making it immediately applicable to mmWave systems.  Expressions were given for a deterministic (fixed) and random geometries.  The work could readily be extended to process other than BPPs, such as Poisson point processes, as well as networks of shapes other than an annulus.  Due to space constraints, only a few examples have been shown to confirm the accuracy of the approach; a more detailed analysis could use these expressions to provide insight into the role of various parameters such as the number of interferers ($K$), array parameters, channel coefficients, and SNR.  While the focus on this paper has been on mmWave, other applications are possible related to cellular networks, distributed MIMO systems, and more elaborate MAC protocols.  For instance, the same methodology could be used to model channel access schemes with various types of collisions (e.g, full and partial) each with their own severity and probability.  
\balance

\small
\bibliographystyle{ieeetr}
\bibliography{./References}
\end{document}